\documentstyle[aip,fullpage,doublespace,12pt,supercite,psfig,epsf]{article}

\def\sqr#1#2{{\vcenter{\hrule height.#2pt
      \hbox{\vrule width.#2pt height#1pt \kern#1pt
        \vrule width.#2pt}
      \hrule height.#2pt}}}

\begin{document}

\bibliographystyle{jcp}

\title{Application of A Distributed Nucleus Approximation In Grid Based
Minimization of
the Kohn-Sham Energy Functional }
\author{Karthik A. Iyer, Michael P. Merrick, and Thomas L. Beck\\
\em{Department of Chemistry}\\
\em{University of Cincinnati}\\
\em{Cincinnati OH, 45221-0172}}

\maketitle

\begin{abstract}
{ In the distributed nucleus approximation  we represent the singular nucleus
as smeared over a small
portion of a Cartesian grid. Delocalizing the nucleus allows us to  solve the
Poisson equation for the
overall electrostatic potential using a linear scaling multigrid algorithm.
This work is done in the context of minimizing the Kohn-Sham energy functional
directly in real space with a multiscale approach. The efficacy of the
approximation is illustrated by
locating the ground state density of simple one electron atoms and molecules
and more complicated multiorbital systems.
}
\end{abstract}

\listoffigures
\newpage
\listoftables
\newpage

\section{Introduction}

The conventional approach to solving electronic structure problems has been
through the use of basis set expansion of wavefunctions.\cite{szabo/ostlund}
While these methods can
produce highly accurate results, there are a few drawbacks. Amongst them, the
completeness of the basis set is always a concern, treating aperiodic systems
with plane wave bases leads to waste in computational effort, and most
importantly the method scales unfavorably with system size.
Several recent studies have shown that accurate $ab-initio$ results can be
generated by a real space representation of the same problem. By decomposing
the multicenter problem into several single center problems and by propagating
the orbital residues in real space, Becke\cite{becke} has obtained impressive
accuracy in
Density Functional calculations on polyatomic systems. More recently,
Chelikowsky et al.\cite{chelikowsky/troullier/saad} have developed a finite
difference-psuedopotential method
and successfully applied it to the $ab-initio$ computation of properties of
several diatomics.

Considerable effort has been expended in recent years towards developing
linear scaling solutions to the electronic structure problem.
Researchers have focussed on using multigrid (MG) methods and/or localized
orbitals to overcome the unfavorable scaling.
Bernholc et al.\cite{bernholc/yi/sullivan} have used a full MG algorithm with
non-uniform grids to
perform real space electronic structure calculations. They present results for
H and H$_2$. Davstad\cite{davstad} has discretized the Hartree-Fock (HF)
equations and used
MG methods to solve the resultant equations for diatomics. Teeter
and coworkers\cite{white/wilkins/teter}  have used a finite element basis in
conjunction with MG to solve
for the electronic structure of several one orbital systems.
 Baroni and Gianozzi\cite{baroni/giannozzi} represented the Hamiltonian in real
space and developed a Lanczos method which solves directly for the
ground state electron
density.
Within the plane wave basis scheme,  Galli and
Parrinello\cite{galli/parrinello} have proposed a nonorthogonal, localized
orbital
approach.
Mauri et al.\cite{mauri/galli/car} and Ordejon et
al.\cite{ordejon/drabold/grumbach/martin} have
developed related methods employing localized, orthogonal
(or generalized Wannier) orbitals.
Stechel et al.\cite{stechel/williams/feibelman} have presented a general
algorithm for iteratively obtaining
the occupied subspace using nonorthogonal, localized orbitals. A
different approach has been taken by three
groups\cite{li/nunes/vanderbilt,daw,carlsson} who have developed
methods for variational solution for the one electron density matrix.
These methods utilize cutoffs in the density matrix beyond some length
scale, and a `purification transformation' to preserve idempotency in
the density matrix. Finally, an exact path integral formulation of
Kohn-Sham (KS)-Density Functional Theory (DFT) has been
developed\cite{harris/pratt,yang,pratt/hoffman/harris} in the last
ten years which is the single approach using only the diagonal one electron
density.

In the Density Functional Theory\cite{parr/yang,dreizler/gross} -Local Density
Approximation (LDA)
, solving for the ground electronic state of a collection of nuclei
and electrons is equivalent to minimization of the Kohn-Sham Energy
Functional (KSEF). In broad terms, the principal components of a real space
minimization of the KSEF
are: (1) solving for the electrostatic
potential due to the nuclei and electrons, which serves as an input for (2)
propagating the KS orbitals while maintaining orthonormality. The evolving KS
orbitals define a new electronic distribution, which in turn defines a new
potential for the orbitals. Several approaches exist for the iteration of the
above process to self-consistency. It is essential that both (1) and (2) be
solved by linear scaling methods to achieve favorable scaling for the entire
solution process.

Orthogonalization of $N$ delocalized orbitals requires $O(N^3)$ steps. In the
context of generalized
Wannier functions\cite{kohn,kohn-II}, one can obtain orbitals that are
exponentially localized in
systems with band gaps and localized as polynomials for metals. One of the
principal advantages of the real space approach is that these localized
orbitals need not be orthogonalized if they possess no overlap in space. If
such is the case, then methods such as Full Approximation Scheme-Multi Grid
(FAS-MG) developed by Brandt et al.\cite{brandt,brandt/mccormick/ruge} can be
used to propagate
the KS orbitals in a rigorously $O(N)$ scheme.

Conceptually, we are then left with the task of generating the electrostatic
potential due to the electrons and nuclei by a linear scaling
method.
 Traditionally, FFT methods (scale
as $NLogN$) have been used to solve for  the potential resulting from the
electron-electron and electron-nuclei interaction.
Becke's\cite{becke/dickson} method generates the potential by
decomposing the charge density around various nuclei in the system.
The Poisson equation is solved on a radial and angular mesh around each
ion center.
The overall potential is recovered by addition of the single center potentials.
The electrostatic energy due to the interaction of nuclei  has
typically been solved by Ewald (scales as $N^{3/2}$) summation.
York and Yang\cite{york/yang} have modified the Ewald method to develop the
fast Fourier
Poisson method that scales as $NLogN$.
We have developed a physically
intuitive method that solves for the entire electrostatic potential
`in one shot' and exhibits rigorous linear scaling.\cite{merrick/iyer/beck} It
involves approximating the singular nucleus as
distributed over a portion of the grid and solving the Poisson equation for
the resultant charge distribution (electrons and nuclei) using
a full multigrid solver. In this research,
we use a unit cube of charge multiplied by the atomic number $Z$.
The size of the cube at a given scale
is dictated by the grid separation $h$.
The electrons and nuclei are thus placed
on an equal footing in terms of the Poisson equation. In this way,
the entire electrostatic problem is solved, including all electron-electron,
electron-nucleus, {\it and} nucleus-nucleus interactions, in a fast
linear scaling step. A distinct advantage of this approach is that
it obviates the need for Ewald summation to compute the
nuclear contributions to the total energy for periodic systems;
we
have computed electrostatic
energies of periodic ionic lattices to high accuracy with this
method.\cite{merrick/iyer/beck}

This work deals with the use  of this novel approach to solve for the
electrostatic
potential in minimization of the KSEF.\cite{acs/dc/note}
 We have also used a simple nested iteration
scheme, as a precursor to incorporating Brandt's FAS-MG, to propagate the KS
orbitals in coordinate space. Section II deals briefly with DFT-LDA and
presents details of our algorithm. In the following section, we present results
on model multi-orbital atomic and molecular systems to exhibit the accuracy of
the distributed
nucleus approximation.
We summarize our findings and discuss
future research plans in Section IV.

\section{Theory and Methods}

\subsection{Definitions}

The Kohn-Sham total
energy\cite{parr/yang,payne/teter/allan/arias/joannopoulos} can be represented
as (we consider only doubly occupied
states here, and atomic units are used throughout):

\eject
\begin{equation}
 E[\{ \psi_i \}] = 2 \sum_{i=1}^{N/2} \int \psi_i^* \left[ - \frac{1}{2}
\right] \nabla^2 \psi_i d^3 {\bf r} + \int V_{ion}({\bf r}) \rho ({\bf r})
d^3 {\bf r}
\label{eq:ksef}
\end{equation}
$$
+ \frac{1}{2} \int \frac{\rho({\bf r}) \rho({\bf r\prime})}
{\mid {\bf r} - {\bf r\prime} \mid } d^3 {\bf r} d^3 {\bf r\prime}
+ E_{XC}[\rho ({\bf r})] + E_{nucleus}(\{ {\bf R}_N \} ) . $$

\noindent
The set of all wavefunctions, $\{ \psi_i \}$, are the occupied one electron
orbitals.
The first term is the total kinetic energy, the second
is
due to the electron-nucleus electrostatic energy, the third is the
electron-electron electrostatic  interaction, the fourth is the
exchange-correlation
energy, which if known exactly would give the exact ground state
energy, and the final term is the total nucleus-nucleus electrostatic
energy.

The electron density is given by:

\begin{equation}
\rho({\bf r}) = 2 \sum_{i=1}^{N/2} \mid \psi_i ({\bf r}) \mid ^2 .
\label{eq:rho}
\end{equation}

\noindent

The objective, then, is to determine the set of KS orbitals,$\{ \psi_i \}$,
that
minimize the Kohn-Sham energy functional. The self-consistent solution of the
KS equations define the orbitals that minimize the KSEF:

\begin{equation} [- \frac{1}{2} \nabla^2 + V_{eff} ] \psi_i({\bf r}) = E_i
\psi_i({\bf r}),
\label{eq:ks}
\end{equation}
where
\begin{equation} {\mbox v}_{eff}({\bf r}) =
{\mbox v}_{ion}({\bf r}) + \int \frac{\rho({\bf r'})}
{\mid {\bf r} - {\bf r'} \mid} d {\bf r'}
+ {\mbox v}_{xc}(\rho({\bf r})) .
\label{eq:terms}
\end{equation}
The first two terms in the effective potential are the total
electrostatic contribution to the electronic part of the
total energy, which is long ranged, while the
exchange correlation potential in the LDA depends only on the local
electron density.

We have used the exchange-correlation potential of Vosko et
al.\cite{vosko/wilk/nusair}
(VWN) which was parametrized from the Monte Carlo data of Ceperley
and Alder.\cite{ceperley/alder} We have assumed the paramagnetic form here
since we are only
interested in doubly occupied states in this work.

\subsection{Grid Representation}

We represent the wavefunctions and operators on an evenly
spaced Cartesian grid. The nuclei are represented as a cube of charge located
at the grid point corresponding to the nucleus position. The effective
potential (operator) is diagonal in the coordinate representation; thus
its application is trivial. In this paper, we represent the
kinetic energy operator using a  finite differences(FD) representation.
 For atomic and molecular
problems, we find that we need at least a  6$^{th}$ order FD form
to obtain accurate results; all computations in this work have used an
8$^{th}$ order form.
Our findings are consistent with
those of Chelikowsky et al.\cite{chelikowsky/troullier/saad}, who
recently discussed use
of a FD representation in DFT calculations.

Full or `exact' solution of the grid problem corresponds to
completely solving
a discretized version of the continuous problem.
 Thus, there are two issues of accuracy. First,
how accurate is the grid representation of the partial differential
equations? Second, how close is one to a complete solution to the
grid represented problem? We note here that, since our problem is not
represented by a Hamiltonian in a complete
basis set, one is $not~guaranteed$ total energies above the exact
ground state. That is, the grid-based approach is a variational
calculation, but does not necessarily satisfy the variational
theorem. One simply knows that by going to a higher resolution
representation, results closer to the exact energy will be
obtained if the problem is completely solved at that finer
scale.

\subsection{Minimization Strategy}

In order to locate the ground state electron density, we must either
solve the Kohn-Sham one electron equations (Eqn. \ref{eq:ks}) to self
consistency, or
(equivalently)
directly minimize the KSEF (Eqn. \ref{eq:ksef}) with respect to wavefunction
variations.   The latter leads to the familiar steepest descent equation:

\begin{equation} \dot{\psi_i}({\bf r}) = -\frac{\delta E[\{\psi_i ({\bf
r})\}]}
{\delta \psi_i^* ({\bf r})} = 0 ,
\label{eq:steep}
\end{equation}

Locating the ground state amounts to propagating Eqn. \ref{eq:steep}  until
a limit in the magnitude of the forces is reached(while
maintaining orthonormality constraints).
We have minimized the KSEF by using steepest descent, Gauss-Seidel (SOR)
and conjugate gradient methods. We have experimented with various `step sizes'
for
steepest descent and Gauss-Seidel calculations and chosen the one that leads
to fastest convergence.
In Gauss-Seidel
propagation, the updated wavefunction value at grid
point $i-1$ is used to update the old value at grid point $i$.
That is, instead of updating all values and then writing
the new wavefunction vector into the old, the new values
are written sequentially as the propagation passes through
the grid.
We found Gauss-Seidel
propagation to be substantially more efficient than steepest descent, and
we employed Gauss-Seidel in our later calculations.
Conjugate gradient provides an efficient and robust minimizer. We have
used the algorithm developed by Payne et
al.\cite{payne/teter/allan/arias/joannopoulos} in certain minimization
calculations. In our method, however, all of the propagation equations are in
coordinate space.

The wavefunctions are orthogonalized at each step of propagation
by the Gram-Schmidt procedure.
The method is
efficient and accurate,
it breaks possible spurious symmetries generated by initial
conditions in the wavefunctions, and leads to the preservation
of ordered energy states.
Orthonormalization is essential
to prevent the collapse of all wavefunctions to the ground state.
With the orthonormalization, the minimization is a well-defined
process for the many electron problem; it should,
when completed, locate
a single minimum in the energy functional represented on the grid
(although multiple minima may in principle occur, we did not encounter
unphysical states in this work). The resulting electron density is the
grid solution to the functional minimization problem of locating
the ground state electron density in Kohn-Sham theory.

\subsection{Nested Iteration for the Orbitals}

In the nested iteration for the wavefunction variational
calculation, minimization is carried out on each
scale until the solution reaches a limit value, beginning
on the coarsest scale. Then, the
solutions (both the wavefunctions and the electrostatic potential)
are linearly interpolated to the next finer
scale, and minimization begins on this scale.
The process is continued until the finest scale is reached, where
we iterate until a self-consistent solution is obtained.
Typically, we use three grid levels, where the
next finer grid spacing was always a factor of two smaller
than the previous coarser scale.

\subsection{Multigrid for the Poisson Equation}

At each iteration step in the minimization process, the Poisson
equation must be solved to generate the electrostatic portion
of the evolving effective potential:

\begin{equation}
\nabla^2 \phi ({\bf r}) = - 4 \pi \rho ({\bf r}).
\label{eq:poisson}
\end{equation}

\noindent
We solve this equation using a full multigrid cycle.
Multigrid for the Poisson equation is known to be
a linear scaling process.
The solution of the Poisson equation is embedded within
the nested iteration for the orbitals.
 The Poisson equation is discretized on the same grid as the
Schr{\"o}dinger equation. For consistency,
the same representation (8$^{th}$ order FD) for $\nabla^2$
is used as for the kinetic energy operator.

The Poisson equation is an elliptic partial differential
equation: solution requires the input of the charge density and
boundary conditions (either finite or periodic). In this paper,
we treat finite systems and fix the values of $\phi({\bf r})$
at the boundaries of the grid.  Once a new value of the orbitals
is obtained following a propagation step, a new charge density is
constructed, the old values of the potential are taken as the
initial $\phi({\bf r})$(except for the very first
solution of the Poisson equation), and the multigrid
process is initiated. Since the input values of $\phi({\bf r})$
are then relatively close to the solution, the process is
rapid.

On a grid, the Poisson equation
can be written as the following matrix equation:

\begin{equation}
{\bf Au} = {\bf b}
\label{eq:matrix}
\end{equation}

\noindent
where $\bf A$ is the matrix representation of the $\nabla^2$ operator
in real space, $\bf u$ is the potential vector on the grid, and $\bf b$ is
the vector representing $- 4 \pi \rho ({\bf r})$. If $\bf u$ is the exact
solution for the given
fine grid representation, and $\bf v$ the evolving solution during
iteration, then Eqn.\ref{eq:matrix} can equally well be represented as:

\begin{equation}
{\bf Ae} = {\bf r}
\label{eq:error}
\end{equation}

\noindent
where ${\bf e} = {\bf u} - {\bf v}$ (the error) and ${\bf r} = {\bf b}
- {\bf Av}$ (the residual). The residual is known at the beginning
of the iteration, and solution of Eqn. \ref{eq:error}  yields the complete
error
in the initial guess for the solution. By adding $\bf e$ to $\bf v$,
the solution $\bf u$ is obtained.
Once the fine grid approximation is obtained, the multigrid
cycle generates corrections to the initial guess by passing
the error and residual to
coarser scales (restriction process) and iterating on the residual
equation. This process is carried out on a sequence of grids going
from fine to coarse and back to fine (interpolation).
The mathematical arguments
for the convergence behavior of multigrid are subtle and are not
discussed here. In words, by passing to a coarser scale, the
long wavelength modes in the error
appear more oscillatory and are thus damped at a greater rate.
By generating corrections on a sequence of coarser scales
and passing this error information back to the fine scale,
critical slowing down
can be overcome. The resulting algorithm is fast and linear
scaling, often requiring on the order of 10 iterations
or less on the fine scale.
Interested readers are referred to
excellent review articles of Brandt\cite{brandt} and Briggs\cite{briggs} for
more details.

\section{Results}

\subsection{Hydrogen-like atoms and H${_2^+}$}
One approach to solve for the relevant orbital is to treat these one electron
systems
as paramagnetic and incorporate the full effective
potential.\cite{puska/nieminen/manninen} Instead, we
treat the electrostatic potential as a function of the bare nucleus alone.
This reduces the problem to a fixed potential eigenvalue problem and provides
a stringent test of the distributed nucleus approximation.
As stated earlier,
the nucleus is represented as a cube of charge, of magnitude Z, at the
grid point corresponding to the nucleus position.
Being a long ranged potential, and with no electronic
density to shield the nucleus, Z/R boundary conditions are imposed
on $\phi({\bf r})$  when
solving the Poisson equation. Since we treat the electrons and nuclei
together, our potential has an additional nucleus self interaction term
associated with it. This self energy is a constant for given Z and grid
spacing and needs to be subtracted from the computed energy to obtain the true
energy of the system.

The energy functional
has been minimized on a three-tiered regular
Cartesian grid. The grids are labelled coarse, intermediate and fine.
We have experimented with several starting configurations for the orbitals
ranging from particle-in-a-box states to hydrogen-like wavefunctions. We
propagate the orbitals using  the steepest descent, Gauss-Seidel  or conjugate
gradient recipes.
 Regardless of the starting configuration we find convergence to be rapid.
Table  \ref{tab:h-like-results} summarizes the
results for these calculations, where we have started with particle-in-a-box
states. We note several interesting points:  We are
pleasantly surprised by the accuracy of the distributed nucleus approximation
in solving for the ground electronic state. We compute accurate ground state
energies and the virial theorem is satisfied to around 1\% accuracy.
As expected, more accurate results
are obtained as we go to finer grids. Even using a simple nested scheme
(as opposed to the FAS-MG) we find great gains in computational efficiency in
locating the minimum of the energy functional.

Table \ref{tab:h2+-results} presents the results for the H${_2^+}$ at various
internuclear
separations. The grids were chosen to accommodate each nucleus on grid
points. This calculation illustrates that we are able to generate accurate
absolute energies and
equally accurate binding energies.\cite{wind} Generation of accurate binding
energies is
crucial in structure determination and Monte Carlo or Molecular
Dynamics calculations.

\subsection{Neon}
We now attempt to locate the ground state electronic density of the 10
electron Ne atom. Our calculation, DFT-LDA (VWN), treats all electrons
explicitly and
involves the propagation of five KS orbitals. Within the DFT-LDA approximation
the energy for the neon atom is
 -128.214 Hartree.\cite{perdew/zunger} The Hartree-Fock energy of the same is
-128.547 Hartree.\cite{fischer}

As with hydrogen-like atoms, the nucleus is treated as a cube of charge.
However, there are a few important differences between this multiorbital
computation  and
the previous one electron calculations. Since the electrostatic potential is
dependent on the charge density, the Poisson equation has to be re-solved with
every update of the wavefunctions.
Further, due to the shielding of the nucleus by the
electronic density we expect the potential to vanish at the boundaries of the
computational grid.

\begin{figure}
\epsfysize=6.0in
\centerline{\epsffile{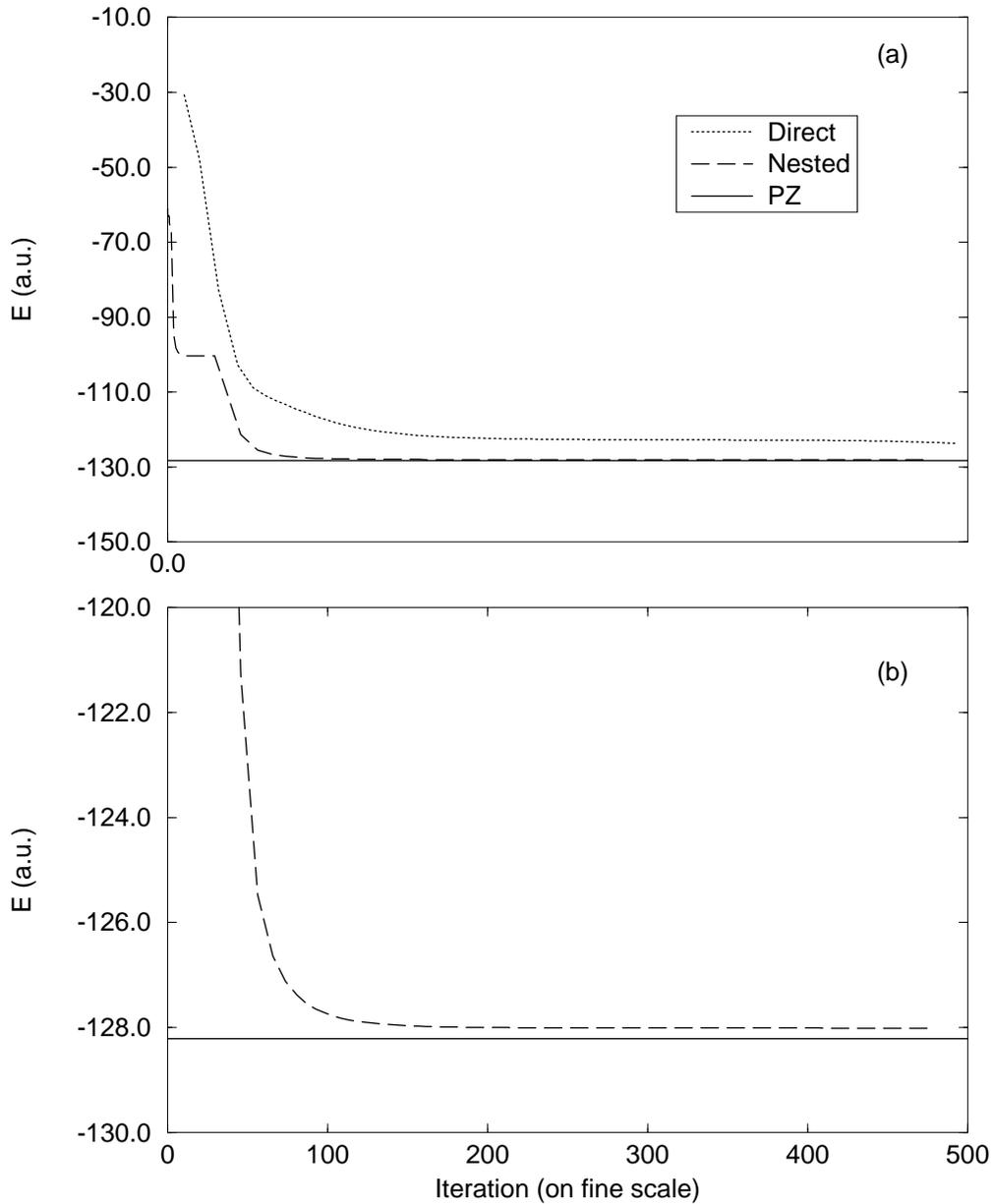}}
\caption{Figure a  illustrates the significant gain in computational
efficiency that one obtains with a multiscale method to propagate the KS
orbitals. For the nested scheme, iterations on the coarse scale are weighted by
a factor of 1/64 and those on the intermediate scale by 1/8. The fact that
critical slowing down is not completely eliminated by the nested scheme is
illustrated in Figure b.}
\label{fig:fig-I}
\end{figure}

Our results for these calculations, where we have started with
particle-in-a-box states, are summarized in  Tables
\ref{tab:ne-results}, \ref{tab:ne-moments-r} and \ref{tab:ne-moments-r2}. As
stated earlier, we are not guaranteed total energies above that of the exact
ground state and this is borne out in the calculation with $h_{fine}$=0.90/4
(Table \ref{tab:ne-results}). However, for a given grid spacing the
minimization is robust and we solve the discretized equations accurately. Using
a nested iteration scheme with $h_{fine}$=0.95/4
we are within 0.5\% of the calculated energy for the neon atom. This result
was obtained by iterating 256 times on the coarse, 128 times on the
intermediate and 64 times on the fine grid.
Figure \ref{fig:fig-I}a illustrates the significant gains one obtains by
using this approach as opposed to iterating only on the fine scale. One
requires on the order of 10$^3$ direct iterations on
the fine scale alone to obtain an energy within 1 Hartree of the converged
results. In contrast, we require merely 20 iterations on the fine scale with
the nested scheme to attain such accuracy. Also note that iterations on the
coarse scale are roughly 1/64$^{th}$ (and 1/8$^{th}$ on the intermediate
scale) as expensive as on the fine scale. In addition, iterating on the coarser
scales
allows us to remove substantial portions of  the long wavelength errors in the
orbitals more
efficiently. While we have reduced the effects of critical slowing down (CSD),
the phenomenon is not completely eliminated (Figure \ref{fig:fig-I}b)  by the
nested scheme. After 64 iterations on the fine scale the energy of the system
is
computed to be -127.900 Hartree. After 128 iterations it is -127.992 Hartree,
after 256 iterations it is -128.011 and finally, after 512 iterations the
energy
for the neon atom is -128.013 Hartree. Thus, a considerable number of expensive
fine scale iterations have to be performed to obtain the final converged
solution. Also, we observe deviations from the
Virial theorem  by as much as  15 to 20\%. Our investigation indicates that
this is primarily due to the poor representation of the core 1s orbital, which
contains an overwhelming portion of the energy is. A calculation for a
hydrogenic atom
with Z=10 (Table \ref{tab:h-like-results}),
 with the same grid as for neon, indicated a similar 15$\%$ departure.

\begin{figure}
\epsfysize=5.0in
\centerline{\epsffile{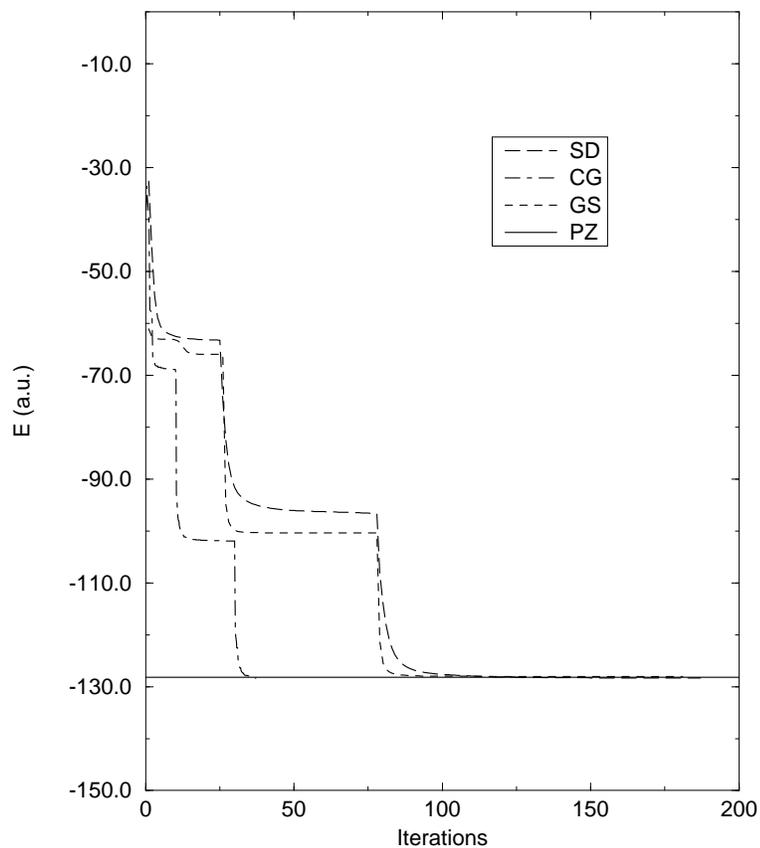}}
\caption{This figure illustrates that some degree of  CSD remains on all
scales, even though the nested cycle has led to substantial acceleration. We
have
made no effort to optimize the number of iterations on each scale. While
conjugate gradient requires the fewest number of iterations, the time per
iteration (update of all five orbitals) is roughly 3-4 times that of
Gauss-Seidel, making the methods competitive.}

\label{fig:fig-II}
\end{figure}

To analyze the convergence of the KS orbitals we have calculated their radial
moments as the solution evolves. $\langle R \rangle$ probes the regions closer
to
the nucleus and $\langle R^2 \rangle$ provides an understanding of
the tails of the orbitals. The $\langle R \rangle$ and $\langle R^2 \rangle$
we calculate are somewhat different from those calculated by Perdew and
Zunger.\cite{perdew/zunger} To facilitate a direct comparison with their
calculation we have used a modified VWN potential with the exchange term
only.\cite{compare}
That our results are different  should come as no surprise since the grid
representation of the orbitals
is somewhat crude. Despite this, we observe atomic shell  structure in the
radial distribution function consistent with HF calculations. It appears that
the convergence of the 1s
orbital to the eventual solution is rapid.
We speculate
that two interlinked factors contribute to the somewhat slower convergence
of the 2s and 2p orbitals. Firstly, since the forces on the 1s orbital are
much greater (due to its proximity to the nucleus) the convergence process
is accelerated. This is borne out by the rate of convergence of $\langle R
\rangle$
and $\langle R^2 \rangle$ for this orbital (no wonder the frozen core
approximation is so good!). The 2s and 2p orbitals are more delocalized
and therefore have longer wavelength modes associated with them.
It is well known that residual modes of $\lambda \leq 4h$ (high frequency) are
readily eliminated by iterating on a  scale with grid spacing $h$. It is the
elimination of the longer wavelength error modes in the evolving solution that
causes CSD (Figure \ref{fig:fig-I}). Thus, while the nested scheme provides
significant improvement
we still encounter some CSD. This appears to be independent of the method
(Figure \ref{fig:fig-II}) chosen to propagate the KS orbitals.

\section{Discussion}

In summary, we have used the distributed nucleus approximation to compute the
overall electrostatic potential accurately with a liner scaling algorithm. We
have obtained encouraging results in using this approximation for the solution
of the Kohn-Sham orbitals for single electron and multiorbital cases. In
general, we compute energies to high accuracy, and the orbital representation
is adequate. We feel that this method can be successfully employed to
perform large scale simulations of interesting condensed phase systems. For the
purposes of high resolution electronic structure calculations one would need
significantly larger number of grid points.

The nested iteration scheme highlights the importance of length scales in
solving for the KS orbitals. We have presented clear evidence that direct
iteration on the fine scale alone is an inefficient process. The use of coarser
scales enables us to obtain dramatic improvements in convergence and postpones
the onset of critical slowing down until we are closer to the eventual
solution. This phenomenon is evident in all three propagation methods and
suggests that smoothing of long wavelength modes of the error in the solution
is of more importance than
the propagation method used on each scale.

The evidence we have presented suggests that we would benefit greatly by
adopting the
FAS-MG scheme. The advantages are: (1) the method scales rigorously in a linear
 fashion as long as
the $N^3$ orthogonalization bottleneck is overcome by use of localized
orbitals and thus, critical slowing down is completely overcome (as has been
done
for in the solution of the Poisson equation), (2) the method lends itself
readily to the use of adaptive grids which should improve the orbital
representation around the nucleus, and finally (3) with the incorporation of
computational `zones of refinement' the storage requirements can be reduced to
 modest amounts in
large scale simulations.

\section{Acknowledgments}
We would like to thank Achi Brandt, David Hoffman, Donald Kouri,
Randall LaViolette, Thomas Marchioro II,  Ruth Pachter, Frank Pinski,
Lawrence Pratt, Ellen
Stechel, and Priya Vashishta for helpful discussions concerning
this work. This research was supported by NSF grant CHE-9225123.
We thank the Ohio Supercomputer Center for their support of this
project through grant time on the Cray-YMP,on which some of these
calculations were performed.
TLB would like to thank AFOSR and Dr.\ Ruth Pachter at Wright
Patterson Air Force Base for support during a summer faculty
fellowship.

\newpage

\begin{table}
\begin{tabular}{|c|c|c|c|c|c|c|}
\hline
\bf Z & \bf n$_{fine}$ & \bf h$_{fine}$ & \bf Nested ? & \bf $\Delta$ E\% & \bf
[1- E$_{pot}$/2E$_{kin}$]\%  & \bf Iterations \\
\hline
    1 & 4 & 1.2  &  No & -4.882 & 4.728 & 35 \\
\hline
    1 & 8 & 0.6  &  No & -0.623 & 2.770 & 57 \\
\hline
    1 & 16& 0.3  &  No &  0.268 & 1.516 & 186\\
\hline
    1 & 16& 1.2/4& Yes &  0.267 & 1.609 & 44 \\
\hline
    1 & 4 & 1.4  &  No & -6.200 & 2.694 & 37 \\
\hline
    1 & 8 & 0.7  &  No & -1.470 & 2.412 & 61 \\
\hline
    1 & 16& 0.35 &  No & -0.096 & 0.637 & 198\\
\hline
    1 & 16& 1.4/4& Yes & -0.096 & 0.577 & 46 \\
\hline
    5 & 16& 0.4/4& Yes & -0.582 & 0.577 & 64    \\
\hline
    10& 16& 0.225/4& Yes & -0.833 & 1.190 & 68   \\
\hline
    10& 16& 0.950/4& Yes & -2.236 & 15.219 & 64   \\
\hline
  100 & 16& 0.03125/4& Yes & -2.092 & 2.521 & 73   \\
\hline
\end{tabular}
\caption{Results for calculations on hydrogen-like atoms using the distributed
nucleus approximation. Z refers to the nuclear charge. There are $2 n_{fine}
+1$ lattice points (in one dimension) on the calculation grid. An answer of Yes
under
$\bf{Nested ?}$ indicated that a three tiered scheme was used to propagate the
KS orbitals. An answer of No corresponds to direct iteration on only the
single grid with grid spacing $h_{fine}$. We specify only the number of
iterations on the finest grid. Note that no effort has been made to optimize
grid spacing for Z=5, 10 and 100. The calculation for Z=10 and $h_{fine} =
0.95$ was performed to compare against calculations for the neon atom.}
\label{tab:h-like-results}
\end{table}

\begin{table}
\begin{tabular}{|c|c|c|c|c|c|c|}
\hline
\bf R &  \bf $\Delta$ E\% & \bf $\Delta$ E$_{binding}$\% & \bf ($\langle
T_{calc} \rangle
- \langle T_V \rangle$)/$\langle T_V \rangle$ \% & \bf ($\langle V_{calc}
\rangle - \langle V_V \rangle$)/$\langle V_V \rangle$ \%  \\
\hline
   0.6  &   0.192  &  -0.551 & -0.532 & -0.186 \\
\hline
   0.8  &   0.457  &  -3.390 & -0.473 & -0.199 \\
\hline
   1.2  &   0.004  & 0.183 & 0.480 & 0.183 \\
\hline
   1.4  &   0.128  & 1.659 & -1.630 & -0.615 \\
\hline
   1.6  &   0.194  & 0.291 & -0.539 & -0.195 \\
\hline
   1.8  &   0.262  & 0.936 & -0.064 & -0.023 \\
\hline
   2.0  &   0.340  & 0.819 & 0.121  & 0.043 \\
\hline
   2.2  &   0.433  & 0.669 & 0.623 & 0.213\\
\hline
   2.6  &   0.664  & 0.144 & 1.600 & 0.572\\
\hline
\end{tabular}
\caption{Presented are the results for the H$^+_2$ ion at various internuclear
separations. All calculations were performed with the nested iteration scheme
using the Gauss-Seidel algorithm for wavefunction propagation. Grid spacings
were chosen to accommodate the nuclei at specific lattice points. Exact
results were obtained from the calculations of Wind. The last two columns
indicate deviations from the molecular Virial theorem.}
\label{tab:h2+-results}
\end{table}

\begin{table}
\begin{tabular}{|c|c|c|c|c|c|c|}
\hline
\bf h$_{fine}$ &  \bf Energy &\bf [1- E$_{pot}$/2E$_{kin}$]\% \\
\hline
   PZ      & -128.214 & 0.0 \\
\hline
   0.90/4  & -129.096 & 14.353 \\
\hline
   0.95/4  & -127.900 & 15.332 \\
\hline
   1.00/4  & -126.607 & 18.496 \\
\hline
   1.05/4  & -125.237 & 20.592 \\
\hline
\end{tabular}
\caption{We present results for the Ne atom for various grid spacings.
All calculations were performed with the nested iteration scheme
using the Gauss-Seidel algorithm for wavefunction propagation. We have started
these calculations with 9 grid points and 256 iterations on the coarse scale.
As we move to finer scales the number of grid points scales up and the number
of iterations scales down by a factor of 2 for each successive scale. In this
and all other captions `PZ' refers to the DFT-LDA results of Perdew and
Zunger. The last column tabulates deviations from the Virial theorem.}
\label{tab:ne-results}
\end{table}

\begin{table}
\begin{tabular}{|c|c|c|c|c|c|c|}
\hline
             &  \bf 1s       & \bf 2s     & \bf 2p \\
\hline
   HF     &  0.158       &   0.892    & 0.965  \\
\hline
   PZ     &  0.159       &   0.906    & 0.990  \\
\hline
   32     &  0.1069      &   0.9205   & 1.1429 \\
\hline
   64     &  0.1061      &   0.8609   & 1.1007 \\
\hline
   128    &  0.1062      &   0.8264   & 1.0790 \\
\hline
   256    &  0.1062      &   0.8128   & 1.0725 \\
\hline
   512    &  0.1062      &   0.8097   & 1.0914  \\
\hline
\end{tabular}
\caption{$\langle R \rangle$ for the Ne orbitals are presented at different
stages of iteration on the fine scale. For example, $\langle R
\rangle_{1s}$=0.1067 after 32 iterations on the fine scale. This calculation
was performed using the nested scheme with
$h_{fine}$=0.95/4. `HF' refers to moments calculated from Hartee-Fock
wavefunctions. To facilitate a direct comparison with `PZ' we have used a
modified VWN potential with exchange only.}
\label{tab:ne-moments-r}
\end{table}

\begin{table}
\begin{tabular}{|c|c|c|c|c|c|c|}
\hline
             &  \bf 1s       & \bf 2s     & \bf 2p \\
\hline
   HF     &  0.034       &   0.967    & 1.229 \\
\hline
   PZ     &  0.034       &   1.005    & 1.326 \\
\hline
   32     &  0.0323      &   1.0736   & 1.6997 \\
\hline
   64     &  0.0319      &   0.9511   & 1.6025 \\
\hline
   128    &  0.0319      &   0.8715   & 1.5465 \\
\hline
   256    &  0.0319      &   0.8363   & 1.5263 \\
\hline
   512    &  0.0319      &  0.8279    & 1.5219 \\
\hline
\end{tabular}
\caption{$\langle R^2 \rangle$ for the Ne orbitals are presented at different
stages of iteration on the fine scale. For example, $\langle R^2
\rangle_{1s}$=0.0321 after 32 iterations on the fine scale. This calculation
was performed using the nested scheme with
$h_{fine}$=0.95/4. }
\label{tab:ne-moments-r2}
\end{table}
\newpage

\newpage

\newpage

\bibliography{ks}

\end{document}